\documentclass[12pt]{article}

\usepackage{epsfig,amsmath,amssymb,axodraw}

\providecommand{\Sbar}{\overline{S}}
\providecommand{\Tbar}{\overline{T}}
\providecommand{\Zbar}{\overline{Z}}
\providecommand{\Wbar}{\overline{W}}
\providecommand{\jbar}{\bar{\jmath}}
\providecommand{\ibar}{\bar{\imath}}

\providecommand{\hbarr}{\bar{h}}

\numberwithin{equation}{section}

\begin{document}

\thispagestyle{empty}
\phantom{}\vspace{-1.5cm}
\rightline{HU-EP-01/31}
\rightline{hep-th/0108220}
\rightline{\today}
\vspace{-0.5truecm}

\begin{center}
  {\bf \LARGE G-Fluxes and Non-Perturbative\\[2mm]
              Stabilisation of Heterotic M-Theory}
\end{center}

\vspace{1.3truecm}
\centerline{Gottfried Curio $^{a,}$\footnote{curio@physik.hu-berlin.de}
and Axel Krause $^{b,}$\footnote{krause@physics.uoc.gr}}
\vspace{.6truecm}

{\em
\centerline{$^a$ Humboldt-Universit\"at zu Berlin,}
\centerline{Institut f\"ur Physik, D-10115 Berlin, Germany}}
\vspace{.3truecm}
{\em
\centerline{$^b$ Physics Department, University of Crete,}
\centerline{71003 Heraklion, Greece}}

\vspace{1.0truecm}
\begin{abstract}

We examine heterotic M-theory compactified on a Calabi-Yau manifold
with an additional parallel M5 brane. The dominant non-perturbative
effect stems from open membrane instantons connecting the M5 with the
boundaries. We derive the four-dimensional low-energy supergravity
potential for this situation including subleading contributions as it
turns out that the leading term vanishes after minimisation. At the
minimum of the potential the M5 gets stabilised at the middle of the
orbifold interval while the vacuum energy is shown to be manifestly
positive. Moreover, induced by the non-trivial running of the Calabi-Yau
volume along the orbifold which is driven by the G-fluxes, we find that
the orbifold-length and the Calabi-Yau volume modulus are stabilised at
values which are related by the G-flux of the visible boundary. Finally
we determine the supersymmetry-breaking scale and the gravitino mass for
this open membrane vacuum.

\end{abstract}

\bigskip \bigskip
\newpage
\pagenumbering{arabic}

\section{Introduction and Summary}
Eleven-dimensional heterotic M-theory \cite{HW1},\cite{HW2} exhibits
two fundamental model-independent moduli. One, the length $R\rho$ of
the orbifold-interval $S^1/\mathbb{Z}_2$, determines the strength of
the string-coupling. The other, which appears upon compactifying the
theory on a further Calabi-Yau threefold (CY) down to four dimensions,
is the CY volume $Vv$. To make $R$ and $V$ dimensionless, we choose
following \cite{MPS}
\begin{equation}
  \rho = \frac{(2\kappa)^{2/9}}{\pi^{14/9}}
       \simeq 0.2 \kappa^{2/9}\; , \qquad
     v = \frac{\pi}{2^{1/3}}\kappa^{4/3}
       \simeq 2.5 \kappa^{4/3} \; .
\end{equation}
Phenomenological considerations of heterotic M-theory with just the
two orbifold fixed-plane boundary sources
\cite{WWarp},\cite{BD1},\cite{BD2},\cite{LOW},\cite{Ben} imply that
\begin{equation}
 R\rho \simeq 15\kappa^{2/9} \simeq \frac{7.5}{M_{GUT}}\; , \qquad
 Vv \simeq 80\kappa^{4/3} \simeq \frac{1}{M_{GUT}^6}\; ,
 \label{PhenVal}
\end{equation}
or $R \simeq 75$, $V \simeq 32$, where $\kappa^{-2/9}\simeq 2M_{GUT}$
denotes the 11-dimensional Planck-scale and $M_{GUT}=3\times 10^{16}$
GeV the grand unification scale. Therefore the orbifold-modulus is
roughly an order of magnitude larger than the generic CY radius. It is,
however, an important feature of adding a further parallel (to the
boundaries) M5-brane that these tight phenomenological constraints on $R$
and $V$ become relaxed due to the extra freedom coming from the M5's
G-flux (see e.g.~\cite{B},\cite{LPT}). In this case it is even possible
to make $R$ large enough such that $R\rho$ approaches its experimental
upper bound of one millimeter in a large extra dimension scenario.
However this extreme case is highly unnatural and implies a hierarchy
problem \cite{CM}.

It is an intrinsic feature of heterotic M-theory that the magnetic
sources for the G-flux which are its two boundaries lead to a
variation of the CY volume along the orbifold direction. If one
considers the theory from its four-dimensional effective point of view
it is therefore necessary to {\em average} the CY volume over the
orbifold-size which introduces a dependence of $V$ on $R$. Moreover, let
us consider the situation with an additional parallel M5 brane located at
the position $x^{11}=x_{M5}$ along the orbifold-interval. This
configuration guarantees that the M5 is compatible with the supersymmetry
of the heterotic M-theory background and does not break it further. We
assume that the M5 is space-time filling in the four external flat
directions and wraps a holomorphic 2-cycle $\Sigma_{M5}$ of the internal
CY space. In this paper we will restrict ourselves to the case of
$h^{(1,1)}=1$, which covers e.g.~the case of the quintic. It means that
$\Sigma_{M5}$ can be expressed in terms of just one basis holomorphic
curve\footnote{For simplicity we will take $\Sigma$ to be isolated, such
that we do not have to integrate over its moduli describing its position
inside the CY threefold.} $\Sigma$ as $\Sigma_{M5}=\beta\Sigma$ with
positive integer expansion coefficient $\beta$. One can understand
$\beta$ as the number of wrappings of $\Sigma_{M5}$ around $\Sigma$.

The M5 induces an additional G-flux through the relation
\begin{equation}
  \int_{\beta\Sigma}\omega_i = \int_{CY_3} \omega_i \wedge G \; ,
\end{equation}
where $\omega_i$, $i=1,\hdots,h^{(1,1)}$ is a basis of harmonic
$(1,1)$ two-forms and $G=\beta[\Sigma]$ is the four-form which is
Poincar\'e-dual to $\Sigma_{M5}=\beta\Sigma$. Through its induced flux,
the M5 has an influence on the $x^{11}$ dependence of the CY volume.
Namely, the Bianchi identity in the presence of the M5 at position
$x^{11}=x_{M5}$ becomes \cite{WWarp},\cite{LOW}
\begin{alignat}{3}
 dG = -\frac{1}{2\sqrt{2}\pi}\left( \frac{\kappa}{4\pi} \right)^{2/3}
      \Big[ &\sum_{i=1,2}(\text{tr} F_{(i)}^2-\frac{1}{2}\text{tr} R^2)
             \delta(x^{11}-x_{(i)})                      \notag \\
            +&8\pi^2\beta[\Sigma]\delta(x^{11}-x_{M5})
      \Big]\wedge dx^{11} \; ,
\end{alignat}
which in turn leads to the following expression for the CY volume (in
units of $v$) as a function of the orbifold coordinate\footnote{The
Heaviside step-function is defined as $\Theta(x\le 0)=0$ and
$\Theta(x>0)=1$.} \cite{WWarp},\cite{CK},\cite{K1}
\begin{equation}
  V(x^{11}) = V_1+\frac{2}{\rho}
              \left( -r_v
 x^{11}+r_{M5}\Theta(x^{11}-x_{M5})(x^{11}-x_{M5})
              \right) \; .
\end{equation}
The parameter $r_v$ is controlled by the G-flux integrated over the CY
at the visible boundary
\begin{equation}
        r_v = -\frac{1}{8\pi}\frac{\rho}{v}
               \left(\frac{\kappa}{4\pi}\right)^{2/3}
               \int_{CY_3}
               \omega\wedge
               \Big(  \text{tr}(F_{(1)}\wedge F_{(1)})
                      -\frac{1}{2}\text{tr}(R\wedge R)
               \Big)   \; ,
   \label{FluxSlope}
\end{equation}
while the parameter $r_{M5}$ describes the G-flux coming from the M5
brane source
\begin{equation}
 r_{M5} = \frac{\pi\rho}{v}\left(\frac{\kappa}{4\pi}\right)^{2/3}
          \beta \int_{CY_3} \omega \wedge [\Sigma ]
\end{equation}
Notice that both $r_v$ and $r_{M5}$ are positive quantities (for
$r_v$ this holds as long as the ``instanton number''
$-\int_{CY_3}\omega\wedge\text{tr}F_1^2$ on the visible boundary
exceeds the one of the hidden boundary).

The rhs of the Bianchi identity must be cohomologically trivial.
Therefore one arrives by integration over the orbifold-coordinate at the
following anomaly cancellation constraint
\begin{equation}
  \sum_{i=1,2}\Big( \text{tr}(F_{(i)}\wedge F_{(i)})
                   -\frac{1}{2}\text{tr}(R\wedge R)
              \Big)
  +8\pi^2\beta[\Sigma] = 0 \; ,
\end{equation}
which holds at the cohomology level. In terms of the G-flux
parameters a further integration over the CY renders this cohomology
condition into an actual flux-equation
\begin{equation}
  r_v+r_h = r_{M5} \; ,
  \label{ACC}
\end{equation}
where $r_h$ gives the G-flux integrated over the hidden boundary
(i.e. it is formally the same as $r_v$ but with $F_{(1)}$ substituted
by $F_{(2)}$).

Because finally we will need the CY volume in the context of the
four-dimensional effective theory, we have to average it over $x^{11}$
between $0$ and $R\rho$ which gives
\begin{equation}
  V = V_1-r(x)R \; , \qquad
  r(x) = r_v-r_{M5}(1-x)^2
  \label{CYVol}
\end{equation}
where we have expressed the M5-brane position $x_{M5}$ through the
dimensionless parameter $x$
\begin{equation}
 x_{M5}=xR\rho
\end{equation}
with $x\in[0,1]$. Thus the magnitude of the slope in the expression
(\ref{CYVol}) of the average CY volume hinges on both the boundary plus
the M5-brane G-flux in an opposing way. Whereas the boundary flux tends
to curve the volume dependence downwards, the M5 flux tends to bend it
upwards. This counterbalance property will show up prominently in our
stabilised solution later on.

The formulation of heterotic M-theory is only known as a perturbative
expansion in $\kappa^{2/3}$ \cite{HW2}. The leading order $\kappa^{2/3}$
terms give rise to the linear dependence of $V(x^{11})$ on the orbifold
coordinate $x^{11}$. Let us therefore now examine for which parameter
values we can trust the linear approximation. Obviously, we can no longer
trust it when the CY volume $V(x^{11})$ becomes negative,
i.e.~unphysical. A way out when this happens would be to go beyond the
linear approximation and use results of the full non-linear treatment of
the supersymmetric warped background geometry. This would give a
manifestly positive quadratic volume thereby eliminating the negative
volume problem \cite{CK},\cite{K1}. Unfortunately, due to the fact that
in this paper, we will need the K\"ahler-potential later on, which
is only known to first nontrivial $\kappa^{2/3}$ order, we have to seek
for stabilisation within the linear approximation framework and therefore
have to check for its validity.

First, it is obvious that the linear approximation should not break
down, i.e.~encounter a negative CY volume, before having reached the M5
coming from the visible boundary. This then imposes the following
parameter constraint ($x_0^{11}$ denotes the position where the volume
might vanish)
\begin{equation}
  x_0^{11} \ge x_{M5}
  \qquad \Leftrightarrow \qquad
  V_1 \ge 2xRr_v  \; .
  \label{Con1}
\end{equation}
Second, we should also make sure that a negative CY volume does not
appear in the second region between the M5 and the hidden boundary at
$x^{11}=R\rho$. In this second region two things can happen. Either
one has a flux-relation
\begin{equation}
  r_{M5} > r_v \; ,
 \label{Con2a}
\end{equation}
which means that $V(x^{11})$ is increasing beyond the M5 and
thus nullifies the negative volume problem for the second region. Or one
could have
\begin{equation}
  r_{M5} \le r_v \; ,
 \label{Con2b}
\end{equation}
which gives a constant or decreasing $V(x^{11})$ beyond the M5. To
guarantee that $V(x^{11})$ in this second case does not become
negative before the hidden boundary is reached means to constrain the
slope of the running volume which is determined by the fluxes. Therefore,
we have to require in addition that
\begin{equation}
  x_0^{11} \ge R\rho
  \qquad \Leftrightarrow \qquad
  V_1 \ge 2R\big( r_v-r_{M5}(1-x) \big) \; .
  \label{Con2}
\end{equation}
To summarise, we have to require either (\ref{Con1}) with
(\ref{Con2a}) or complementary (\ref{Con2b}) together with
(\ref{Con2}) (notice that (\ref{Con1}) is implied by
(\ref{Con2b}) and (\ref{Con2})) in order to trust the first order linear
volume approximation .

Since the succesful prediction of four-dimensional data, in particular
Newton's Constant \cite{WWarp},\cite{CK}, hinges on the above values
(\ref{PhenVal}), the question arises of how to stabilise them. This
will be the main concern of this paper. There are various
non-perturbative effects which give rise to interesting potentials for
$R$ and $V$. In the framework of the heterotic string the main
non-perturbative mechanism for breaking supersymmetry has been gaugino
condensation in a hidden sector \cite{DRSW}. In the context of
heterotic M-theory gaugino condensation appears even more naturally as
the gauge theory on the hidden boundary now becomes strongly coupled
\cite{WWarp}. Moreover, with the geometrical separation of the two $E_8$
gauge groups there appears yet another class of non-perturbative objects.
These are the open membranes (OM) which either connect one boundary with
the other or with some intermediate M5-brane placed parallel to the
boundaries along the orbifold-interval. Furthermore, also M5-instantons
and M2-instantons can appear. The former wrap the whole internal CY
whereas the latter wrap a 3-cycle of the CY.

In \cite{CKK} it was argued that through the combined effect of
multi-gaugino condensation on the hidden wall together with parallel
(to the boundaries) M2-instantons a phenomenologically satisfactory
stabilisation of the $R$ and $V$ moduli could be achieved. While the
parallel M2-instanton breaks all supersymmetry explicitly \cite{LLO}
and one cannot use supersymmetric tools to derive the potential, other
non-perturbative sources like the mentioned orthogonal (to the
 boundaries) OM's
or M5-instantons are compatible with the supersymmetry of heterotic
M-theory. They will break supersymmetry spontaneously.

In general there are two different stabilisation scenarios which have
to be distinguished. They differ in the energy-scale at which
stabilisation might occur. Either the theory could become stabilised
above the threshold given by the inverse orbifold-size $1/R\rho =
M_{GUT}/7.5$ or below. In the former case one would have to work with the
eleven-dimensional formulation of heterotic M-theory if stabilisation
even trespasses the CY compactification scale $M_{GUT}$ or otherwise with
the effective five-dimensional action \cite{LOSW} between the two
thresholds. This case offers the intriguing possibility that
local supersymmetry gets broken via gaugino condensation
\cite{H},\cite{OLW1} only if energies become so low that the orbifold
 interval
shrinks to a point. However it leads to the phenomenologically
unsatisfactory situation that the mass of the gravitino
\begin{equation*}
  m^2_{3/2} = M_{Pl}^2 e^{K/2} |W|
\end{equation*}
which is proportional to $\Lambda^3_{GC}$ becomes too high. (For a
discussion of this case with an inverse orbifold-length at the
intermediate scale $10^{12}$ GeV see \cite{AQ}).

Therefore, subsequently we will search for a stabilisation in the
energy-regime below the $M_{GUT}/7.5$ threshold, which necessitates a
description of heterotic M-theory through its effective
four-dimensional N=1 supergravity action. This had been derived in
\cite{OLW2}. In particular we will analyse the case of vanishing charged
scalar vacuum expectation values (vev's).

As it turns out that in the regime where one can trust the perturbative
formulation of the effective four-dimensional heterotic M-theory the
non-perturbative M5-instantons and gaugino condensation appear
exponentially suppressed, we will focus on the effect of OM-instantons in
the presence of a parallel M5-brane which is the dominant one.
By minimising the corresponding potential for the moduli, we find that
OM-instantons do stabilise the M5 in the middle of the orbifold
interval. Furthermore, the moduli $V$ and $R$ get stabilised at
values
\begin{equation}
V = \frac{V_1}{4} \;, \qquad  R = \frac{V_1}{r_v} \; .
\end{equation}
To find this minimum of the effective potential it is essential to have
nontrivial G-fluxes caused by the boundaries and the M5. They trigger a
dependence of $V$ on $R$ which is responsible for the stabilisation.
Indeed, for consistency with the perturbative formulation of the theory,
the G-fluxes integrated over the visible boundary and the M5 have to be
equal \begin{equation}   r_v = r_{M5} \; .
  \label{FluxEquality}
\end{equation}
In the full eleven-dimensional picture this OM-instanton vacuum
corresponds to a CY volume which falls off linearly and approaches zero
in the middle of the interval where the M5 is located. For the second
half of the interval it stays constant due to the flux-equality
(\ref{FluxEquality})
\begin{equation}
     V(x^{11}) =
     \left\{ \begin{array}{lll}
     V_1-\frac{2}{\rho}r_vx^{11} & \; , &
     0 \le x^{11}< \frac{R\rho}{2} \\
     \qquad 0           & \; , &
     \frac{R\rho}{2}\le x^{11} \le R\rho
     \end{array} \right.
\end{equation}
We will however show that there is evidence that the full theory
beyond the first order shifts the volume-zero on the second half-interval
to a non-vanishing positive constant value. It is intriguing to see that
the {\em relationship between $R$ and $V$ is simply determined by the
flux $r_v$ coming from the visible boundary}
\begin{equation}
  R = 4\frac{V}{r_v} \; .
\end{equation}
Moreover, since at the minimum the leading order terms vanish it is
important to include all first order $\kappa^{2/3}$ corrections. This
gives a manifest positive contribution to the vacuum energy possessing
an interesting exponential suppression factor.

In the following table\footnote{Here we have chosen a CY-intersection
number $d =30$ and $|h|=\beta=1$. The meaning of $h$ will become clear in
the next section.} we give a quick impression of what will be the
relevant data with OM-instantons present for integrated G-fluxes $r_v$
and CY volumes $V_1$ on the visible boundary and what will be their
respective influence on $R$, $V$, on the two heterotic M-theory expansion
parameters $\epsilon$, $\epsilon_R$, on the vacuum energy $U_{OM}$, on
the related supersymmetry-breaking scale $M_{Susy}$ and finally on the
gravitino mass $m_{3/2}$. A small $r_v$ can be seen to ruin
the smallness of $\epsilon$ and thereby the reliability of the
perturbative formulation of the theory. Thus the $r_v$ fluxes have to be
considerable. At the same time a not too small $r_v$ allows to bring the
supersymmetry-breaking scale into the desired TeV region, however,
simultaneously rises the vacuum energy $U_{OM}$. It can be seen that an
increasing $V_1$ has a similar effect on $U_{OM}$, $M_{Susy}$ and
$m_{3/2}$ as a decreasing $r_v$. However, its influence on $\epsilon$,
$\epsilon_R$ is rather modest. Generically, one obtains a $V$ which is
one or two magnitudes larger than $R$. The basic reason for this is to
keep the parameter $\epsilon$ small enough.
\begin{center}
\begin{tabular}{|c|c||c|c|c|c|c|c|c|} \hline
  $r_v$ & $V_1$ & $R$ & $V$ & $\epsilon$ & $\epsilon_R$ &
  $U_{OM}^{1/4}/\text{TeV}$ & $M_{Susy}/\text{TeV}$ &
  $m_{3/2}/\text{TeV}$ \\
\hline\hline
  90 & 3000 & 33 & 750 & 0.8 & 0.1 & $10^{-5}$ & $3\times 10^{-4}$ &
  $2\times 10^{-6}$ \\
\hline
  140 & 3000 & 21 & 750 & $0.52$ & $0.18$ & 91 & 1815 & 16 \\
\hline
  200 & 3000 & 15 & 750 & 0.36 & 0.25 & $5\times 10^5$ & $9\times 10^6$ &

  $10^5$ \\
\hline
  200 & 4000 & 20 & 1000 & 0.4 & 0.2 & 39 & 782 & 6.7 \\
\hline
  200 & 5000 & 25 & 1250 & 0.4 & 0.2 & $10^{-3}$ & $3\times 10^{-2}$ &
  $2\times 10^{-4}$ \\
\hline
\end{tabular}
\end{center}

The organisation of the paper is as follows. After presenting
preparatory material and the relevant K\"ahler- and superpotential in
section 2, we will derive in section 3 the four-dimensional N=1
supergravity potential for the OM-instanton background. It turns out to
be positive. Its minimisation results in a minimum for which the leading
order terms vanish and subleading terms become important. The M5 gets
stabilised in the middle of the orbifold-interval and $V,R$ obtain values
depending on $V_1$ and the G-fluxes related to the visible boundary and
the M5. In section 4 we analyse the constraints coming from the
perturbative formulation of the theory and show that they require a
flux-equality between those G-fluxes arising from the visible boundary
and the M5. Moreover, we present the eleven-dimensional picture of the
vacuum solution and give its vacuum energy. The final section 5 treats
the issue of supersymmetry-breaking. We derive the supersymmetry-breaking
scale and gravitino mass for the OM-vacuum studied before and compare the
supersymmetry-breaking scale with its vacuum energy. Technical
details related to the derivation of the potential and the determination
of the supersymmetry-breaking scale plus gravitino mass appear in
appendix A and B.

\section{The Effective D=4 Potential}
In the framework of the low-energy four-dimensional N=1 supergravity
description, the moduli potential is obtained from the K\"ahler- and
the superpotential by means of the general formula
\begin{equation}
  (\kappa_4)^4 U = e^K\left( K^{i{\bar\jmath}}D_iW D_{{\bar\jmath}}
                            \Wbar - 3W\Wbar
                    \right)
                +U_D \; ,
    \label{Potential}
\end{equation}
where $D_i W = \partial_i W +K_i\, W$ denotes the K\"ahler-covariant
derivatives, $K_i \equiv \partial_i K$ and
$U_D\sim\sum_a(\overline{C}T^aC)^2$ denotes the D-term contribution. The
index $i$ runs over all moduli. Note that we multiplied the potential $U$
which has mass-dimension four by a factor $(\kappa_4)^4=1/M_{Pl}^4$
to render the right-hand-side of (\ref{Potential}) and thereby $W$
dimensionless. This is done to get rid of various onerous dimensionful
powers of $v$, $\rho$. For consistency we will also choose the
moduli-fields dimensionless in the following.

Thus we need to know the superpotential $W$ and the
K\"ahler-potential $K$. Besides the perturbative trilinear
superpotential
\begin{equation}
W_{(p)} = \lambda_{IJK} C^I C^J C^K \; ,
\end{equation}
where $\lambda_{IJK}$ denotes the Yukawa-couplings there are various
non-perturbative contributions. Recently, there appeared a detailed
analysis of the contributions of open membrane instantons to the
superpotential \cite{MPS},\cite{LOPR}. Either the open membranes connect
both boundaries with each other and give rise to a superpotential
$W_{(M2)}$ or they connect the boundaries with the additional M5-brane
located along the orbifold-interval giving a superpotential
$W_{(M2,M5)}$. In the latter case in order to have a supersymmetric
configuration, the open membrane must have the geometry $\Sigma \times
I$, where $I$ describes the interval in the orbifold direction and
$\Sigma$ denotes the same basis holomorphic curve on which also the M5 is
wrapped. The superpotential is then given by
\cite{MPS},\cite{OLW2},\cite{LOPR}
\begin{equation}
W = W_{(p)}+W_{(M2,M5)}+W_{(M2)} \; ,
\end{equation}
with
\begin{equation}
  W_{(M2,M5)} = h\left( e^{-Z}+e^{Z-\beta T}
                \right) \; , \qquad
  W_{(M2)} = h' e^{-\beta T} \; ,
\end{equation}
where the dimensionless complex prefactors $h,h'$ are related to the
complex structure moduli \cite{MPS},\cite{LOPR}. We will not need their
explicit expressions in the following.

The complex moduli fields are defined by
\begin{equation}
  S = V+\beta Jx^2+i\sigma \; , \qquad
  T = J+i\chi \; , \qquad
  Z = \beta Jx+i\alpha \; ,
\end{equation}
where
\begin{equation}
  J = Ra \; ,
\end{equation}
with $a=(6V/d)^{1/3}$ the K\"ahler-modulus of the CY and
$d=\frac{1}{v}\int_{CY}\omega_1^3$ the CY-intersection number with
$\omega_1$ the basis-element of harmonic (1,1)-forms (remember that
we chose a CY with $h^{(1,1)}=1$ such that the K\"ahler-form reads
$\omega = a \omega_1$). The axions $\sigma$ and $\chi$ arise from two
different components of the eleven-dimensional 3-form
potential\footnote{$A,B=0,\hdots,9$; $\mu,\nu,\rho,\lambda = 0,\hdots,3$;
(anti-)holomorphic CY-indices $\bar{m},m=1,\hdots,3$.} $C_{AB11}$ with
one index tangent to the orbifold. $\sigma$ is dual to $C_{\mu\nu 11}$
\begin{equation}
3V^2\partial_{[\mu}C_{\nu\rho]11}=\epsilon_{\mu\nu\rho\lambda}
\partial^\lambda\sigma
\end{equation}
while $\chi$ comes from
\begin{equation}
C_{m\bar{m}11}=\chi \omega_{1,m\bar{m}} \; .
\end{equation}
The axion $\alpha$ is a combination of $\chi$ and a scalar $\cal{A}$
coming from the KK reduction of the M5's 2-form potential $A^{(2)}$
\cite{MPS}
\begin{equation}
\alpha = \beta(x\chi -a\cal{A}) \; .
\end{equation}
More precisely, if $f$ denotes the holomorphic embedding of the curve
$\Sigma$ into the CY, then $\cal{A}$ arises from the KK decomposition
$A^{(2)}=\pi\rho{\cal A} f^\star (\omega)$ with $f^\star (\omega)$
the pullback of $\omega$ to the cycle.

Geometrically $\beta J$ gives the average volume occupied by an OM
stretching from boundary to boundary while $\beta J x$ resp.~$\beta J
(1-x)$ give the average volume of an OM connecting the M5 with the
visible resp.~hidden boundary. In $S$ we included the higher-order
correction $\beta Jx^2$ which had been found in \cite{MPS}.

Next we have to specify the K\"ahler-potential $K$, which is composed out
of five pieces \cite{MPS},\cite{OLW2},\cite{DS}
\begin{equation}
  K = K_{(S,M5)} + K_{(T)} + K_{(C)} + K_{(cx)} + K_{(bd)} \; ,
\end{equation}
where
\begin{alignat}{3}
  &K_{(S,M5)} = -\ln \left( S+\Sbar
                           -\frac{(Z+\Zbar)^2}{\beta (T+\Tbar)}
                     \right) \; , \qquad\qquad\qquad
   K_{(T)} = -\ln (\frac{d}{6}(T+\Tbar)^3) \label{KSM5T} \; , \\
  &K_{(C)} = \left( \frac{3}{T+\Tbar}
               +\frac{2\xi}{S+\Sbar}
             \right)
             H_{IJ}C^I {\overline{C}}^J+{\cal O}(C^3) \; , \qquad
   K_{(cx)} = -\ln(\overline{\Pi^a}{\cal G}_a) \; ,
\end{alignat}
and the precise meaning of $\xi,H_{IJ},\Pi^a$ or ${\cal G}_a$ can be
found in \cite{MPS}. Unfortunately little is known about the
K\"ahler-potential $K_{(bd)}$ of the instanton gauge bundle
moduli. It has to be noted that this K\"ahler-potential is only valid in
a region where the two heterotic M-theory expansion parameters
\begin{equation}
\epsilon = \frac{2R}{V^{2/3}} \simeq \frac{J}{V}\; , \qquad
\epsilon_R = \sqrt{\frac{\pi}{2}}\frac{V^{1/6}}{R}
             \simeq \frac{\sqrt{V}}{J}
\end{equation}
are smaller than one. Note also that the K\"ahler-potential for the
M5-brane moduli \cite{MPS}, \cite{DS}
\begin{equation}
  K_{(M5)} = \frac{(Z+\Zbar)^2}{(S+\Sbar)\beta (T+\Tbar)}
\end{equation}
is of subleading order $\epsilon$ relative to the leading piece
$K_{(S)}=-\ln ( S+\Sbar )$. It was shown in \cite{DS} that because of
supersymmetry they should be combined into the single
expression $K_{(S,M5)}$ appearing above.

In order to have a well-defined perturbative formulation of the theory,
we have to require that $\epsilon\ll 1$, $\epsilon_R\ll 1$ which means
that we have to restrict ourselves to the region of moduli space
where
\begin{equation}
J^2 \gg V \gg J \gg 1 \; .
\end{equation}
This is the reason why we suppress gaugino condensation and M5-instanton
effects in the present work. Schematically their contribution to the
potential will be exponentially suppressed by $e^{-c_1 V}$ whereas
OM-instantons will exhibit a milder $e^{-c_2 J}$ suppression and hence
dominate ($c_1,c_2$ are positive constants).

In order to gain a better understanding in which region of parameter
space $\{d,r(x),V_1\}$ we obtain small $\epsilon$ and small
$\epsilon_R$, we show some representative values in the following
table

\begin{center}
\begin{tabular}{|c|c|c||c|c|c|c|}
\hline
$d$ & $r(x)$ & $V_1$ & $V$ & $J$ & $\epsilon$ & $\epsilon_R$\\
\hline\hline
  30 & 10 & 500 & 100 & 108.6 & 3.7 & 0.07\\
  30 & 10 & 1000 & 600 & 197.3 & 1.1 & 0.09\\
  30 & 10 & 2000 & 1600 & 273.6 & 0.6 & 0.11\\
  30 & 10 & 5000 & 4600 & 389 & 0.3 & 0.13\\
\hline
  30 & 10 & 5500 & 5100 & 402.7 & 0.3 & 0.13\\
  30 & 50 & 5500 & 3500 & 355.2 & 0.4 & 0.12\\
  30 & 100 & 5500 & 1500 & 267.8 & 0.6 & 0.11\\
  30 & 130 & 5500 & 300 & 156.6 & 1.8 & 0.08\\
\hline
  0.06 & 10 & 2000 & 1600 & 2171.5 & 0.6 & 0.11\\
  0.6 & 10 & 2000 & 1600 & 1007.9 & 0.6 & 0.11\\
  6 & 10 & 2000 & 1600 & 467.8 & 0.6 & 0.11\\
  60 & 10 & 2000 & 1600 & 217.2 & 0.6 & 0.11\\
\hline
\end{tabular}
\end{center}

Here, we kept $R$ fixed (at $R=40$) since it will be determined dynamically
subsequently by minimizing the potentials whereas $\{d,r(x),V_1\}$
are regarded as free ``input'' parameters. From the table it can be
seen that an increasing $V_1$ yields a decreasing $\epsilon$ and a
slightly increasing $\epsilon_R$. On the other hand an increasing
$r(x)$ yields the reversed effect. Finally a varying $d$ has no
influence on $\epsilon$ and $\epsilon_R$ and merely affects the modulus
$J$ which grows when $d$ decreases. Therefore, we will assume $d$
to be fixed at a value of 30 in the rest of this paper. In conclusion
we should look for stabilisation in the parameter-region where $V_1$ is
rather large, say $V_1\gtrsim 2000$ while $r(x)$ should not be too big,
say $r(x)\lesssim 100$.

In this paper we will examine the region of moduli space where charged
scalar $C^I$ (which originate from the reduction of the
ten-dimensional gauge-field) vev's are absent or comparatively
small. Basically this means that we look for stabilisation of
heterotic M-theory at energies high enough such that the GUT gauge
group is still (spontaneously) unbroken. For vanishing $C^I$ the
perturbative contribution $W_{(p)}$ to the superpotential and the charged
scalar K\"ahler-potential $K_{(C)}$ can be neglected subsequently
\begin{equation}
  C^I=0 \qquad \rightarrow \qquad
  W_{(p)} = 0 \; , \qquad K_{(C)} = 0\; .
\end{equation}
It is important to note that in the case with $C^I=0$ the sum
$K_{(S,M5)}+K_{(T)}$ as given by (\ref{KSM5T}) includes all
corrections of order $\epsilon$ and order $\epsilon_R$ (see
e.g.~\cite{Munoz}). This is due to the fact that the subleading
contributions to the leading order expressions for $K_{(S)}$ and
$K_{(T)}$ are proportional to $C^IC^J$ and therefore vanish, while
$K_{(M5)}$ is already of order $\epsilon$.

Moreover for the case of $h^{1,1}=1$ it is known that $W_{(M2)}$
vanishes \cite{VMC},\cite{MPS}
\begin{equation}
  h^{1,1}=1 \qquad \rightarrow \qquad
  W_{(M2)} = 0 \; .
\end{equation}
The moduli-potential for the $h^{1,1}=1, C^I=0$ case which originates
from the contributions \begin{alignat}{3}
 W &= W_{(M2,M5)} \\
 K &= K_{(S)} + K_{(T)} + K_{(cx)} + K_{(bd)}
\end{alignat}
has been first calculated in \cite{MPS} and contains the following first
and next-leading order terms
\begin{alignat}{3}
    (\kappa_4)^4 U_{OM}^{(MPS)}
  = e^{K_{(cx)}+K_{(bd)}}\frac{3|h|^2}{4d J^2}
        &\bigg( e^{-2Jx}+e^{-2J(1-x)}
               -2e^{-J}\cos(2\alpha-\chi)
                                                \label{MoorePotential} \\
              &+\frac{2J}{3V}(1-2x)e^{-2J(1-x)}
               +\frac{4Jx}{3V}e^{-J}\cos(2\alpha-\chi)
         \bigg)+\ldots                          \nonumber
\end{alignat}
Note that here where the $K_{(M5)}$ contribution has not been included
the subleading terms in the second line can give a negative contribution.
We will show in the next section that there are also $x^2$ terms at
subleading order resulting from the inclusion of $K_{(M5)}$ rendering the
potential manifestly positive. Moreover we will see that the leading
order potential vanishes at its minimum and therefore carefully including
all subleading contributions becomes essential.

\section{The Moduli-Potential with OM-Instantons}
Let us now extract the moduli-potential to subleading order resulting
from the two OM instantons connecting the intermediate M5 with either
boundary. The superpotential is given by \cite{MPS}
\begin{equation}
 W = W_{(M2,M5)} = h(e^{-Z}+e^{Z-\beta T})
      \label{InitSupo}
\end{equation}
while the K\"ahler-potential up to subleading order reads
\begin{equation}
 K = K_{(S,M5)} + K_{(T)}
   = -\ln( \frac{8}{3}d V J^3 )
      + {\cal O}(\epsilon^2, \epsilon_R^2,\epsilon\epsilon_R)  \; .
\end{equation}
Notice the inclusion of the subleading $K_{(M5)}$ part. In the expression
for the four-dimensional supergravity-potential (\ref{Potential}) we will
consider only the covariant derivatives $D_iW$ with respect to the
$i=S,T,Z$ moduli, i.e.~we will neglect the dependence on
complex-structure and bundle-moduli.

The potential is composed out of four structurally different parts
which are hierarchically ordered in the $J^2 \gg V \gg J \gg 1$
region. We will examine now their order of magnitude in this moduli space
region. From the superpotential one easily recognizes that partial
derivatives of $W$ with respect to the moduli fields do not generate
further factors of $V$ or $J$. Thus for a determination of the magnitude
of the four different parts, it is sufficient to take the leading
behaviour of the K\"ahler-potential and its derivatives. This can be
found in appendix A.

Let us start with the
\begin{equation*}
   |W|^2
\end{equation*}
term which is of ${\cal O}(1)$ with respect to a counting of
$V$ and $J$ prefactors. The second sort of contribution to the
potential is of the form
\begin{equation*}
  K^{i\jbar}K_iWK_{\jbar}\Wbar \; .
\end{equation*}
The expressions from the appendix show that these terms range between
${\cal O}(1)$ and ${\cal O}(J/V)$ in magnitude. A third class consists of
mixed terms and is given by
\begin{equation*}
  K^{i\jbar}\partial_iWK_{\jbar}\Wbar
\end{equation*}
with ranges between ${\cal O}(V)$, ${\cal O}(J)$ and ${\cal O}(J^2/V)$.
Finally, the last class of terms is
\begin{equation*}
  K^{i\jbar}\partial_iW\partial_{\jbar}\Wbar \; .
\end{equation*}
This class dominates the three others since it exclusively gives the
leading ${\cal O}(JV)$ and subleading ${\cal O}(J^2)$ contributions.
Therefore it is enough to consider just this class to ensure that all
leading and subleading contributions in $\epsilon$ and $\epsilon_R$ are
taken into account. Notice that beyond this order the K\"ahler-potential
of the theory is not known and therefore it would make no sense to
include an incomplete set of terms at these lower orders coming from the
three other classes of terms.

Concerning the last class of dominant terms, the ${\cal O}(J^2)$
contributions which come from $K^{T\Tbar},K^{T\Zbar}$ are suppressed by
$\epsilon$ against the leading ${\cal O}(JV)$ contribution from
$K^{Z\Zbar}$. This means that we have to include for
the latter also its subleading corrections whereas for the former it
is enough to consider merely their leading $\epsilon,\epsilon_R$
behaviour.

\subsection{The OM Potential}
With help of the expressions collected in appendix A one derives the OM
potential for the moduli. {\em Including all leading and subleading $J/V$
corrections} it reads
\begin{alignat}{3}
    (\kappa_4)^4 U_{OM}
 = \frac{3|h|^2}{4d J^2}
   \Bigg\{
    &\beta \Big[ e^{-2J\beta x}+e^{-2J\beta(1-x)}
                -2e^{-J\beta}\cos (2\alpha-\beta\chi)
           \Big]
   \label{OMPotential}                                         \\
   +&\frac{2}{3}\frac{J}{V}\beta^2
           \Big[ e^{-2J\beta x} x^2 + e^{-2J\beta(1-x)} (1-x)^2
                +2x(1-x)e^{-J\beta}\cos (2\alpha-\beta\chi)
           \Big]
   \Bigg\} \; .                                                \notag
\end{alignat}
The symmetry of the potential under the exchange $x\rightarrow 1-x$
originates from the symmetry of the OM-superpotential $W_{(M2,M5)}$
under the exchange of the corresponding moduli $Z\rightarrow\beta T-Z$
(the K\"ahler-potential is trivially symmetric as it does not
depend on $x$). It is important to notice the sign-difference of
the cosine term between leading and subleading order. It is this
difference which prohibits $U_{OM}$ from becoming zero at its minimum
and thereby leads to a {\em spontaneous breaking of supersymmetry}.

An immediate consequence is that this potential is bounded from below
by a non-negative expression
\begin{equation}
   (\kappa_4)^4 U_{OM}
 > \frac{3|h|^2}{4 d J^2}
   \Bigg\{
     \beta \Big[ e^{-J\beta x}-e^{-J\beta(1-x)} \Big]^2
   + \frac{2}{3}\frac{J}{V}\beta^2
           \Big[ e^{-J\beta x} x - e^{-J\beta(1-x)} (1-x)
           \Big]^2
   \Bigg\} \; .
\end{equation}
Since this lower bound can never be saturated, $U_{OM}$ has to be
positive. Hence D=4, N=1 supersymmetry will be broken with a {\em
positive vacuum energy}.

\subsection{Minimisation}
Let us now minimise $U_{OM}$. Minimisation with respect to the axion
fields leads to $\sin(2\alpha-\beta\chi) = 0$ which is solved by
\begin{equation}
2\alpha-\beta\chi = n\pi \; ; \; n \in \mathbb{Z}
\label{Lat1}
\end{equation}
and gives for the potential
\begin{alignat}{3}
  (\kappa_4)^4 U_{OM}
 = \frac{3|h|^2}{4 d J^2}
   &\Bigg\{
     \beta \Big[ e^{-J\beta x} + (-1)^{n+1} e^{-J\beta(1-x)} \Big]^2 \notag\\
   &+\frac{2}{3}\frac{J}{V}\beta^2
     \Big[ e^{-J\beta x} x + (-1)^n e^{-J\beta(1-x)} (1-x) \Big]^2
   \Bigg\} \; .
\end{alignat}
The sectors with
\begin{equation}
     n \in 2\,\mathbb{Z}
\end{equation}
result in a lower energy for the leading term and will be analysed
subsequently. They give the manifestly positive expression
\begin{equation}
  (\kappa_4)^4 U_{OM}
 = \frac{3|h|^2}{4 d J^2}
   \Bigg\{
     \beta \Big[ e^{-J\beta x}-e^{-J\beta(1-x)} \Big]^2
    +\frac{2}{3}\frac{J}{V}\beta^2
     \Big[ e^{-J\beta x} x + e^{-J\beta(1-x)} (1-x) \Big]^2
   \Bigg\} \; .
  \label{Reihe}
\end{equation}
Furthermore, it is easy to see that the value
\begin{equation}
  x = \frac{1}{2}
  \label{Lat2}
\end{equation}
for the M5-brane position modulus $x$ minimizes $U_{OM}$. Hence, the
{\em parallel M5 becomes stabilised at the symmetric position in the
middle of the orbifold-interval}. This could have been anticipated since
both the K\"ahler-potential and the OM superpotential are invariant under
the symmetry which exchanges $x\leftrightarrow 1-x$. Thus, the
OM-potential is mirror-symmetric with respect to the fixed-point $x=1/2$
which means that it must exhibit a minimum or a maximum at the
fixed-point. The explicit analysis confirms a minimum.

It is important to realize that for this value the leading-order part of
the potential vanishes and it is the sub-leading term which contributes
alone and hence becomes responsible for supersymmetry-breaking and a
non-vanishing vacuum energy.

As an aside let us compare our result with the expression
(\ref{MoorePotential}) of \cite{MPS}. The difference lies in the
additional $x^2$ terms which we have included in the subleading terms
and lead to the complete squares. Their origin can be traced back to
the subleading corrections coming from $K^{Z\Zbar}$. We therefore
conclude that it is {\em important to include the
contribution from $K_{(M5)}$ to the K\"ahler-potential} which gives
rise to $K^{Z\Zbar}$ and which seemingly had been omitted in the
derivation of the potential in \cite{MPS}. Finally,
one could be inclined to view (\ref{Reihe}) as the begin of a series
expansion which roughly could be summed up to $\beta e^{-J\beta +
\sqrt{\beta J/V}x}$. This then suggests that higher order in $J/V$
contributions could not endanger the leading-order result when summed
up as long as $J/V \ll 1$.

Before proceeding with the minimisation analysis let us briefly reflect
on a consequence of (\ref{Lat1}) and (\ref{Lat2}). With the definition of
$\alpha$ inserted into (\ref{Lat1}) and setting $x=1/2$, we see that the
axion $\chi$ cancels out and the minimisation condition (\ref{Lat1})
implies setting the scalar $\cal{A}$ to
\begin{equation}
{\cal A} = n\frac{\pi}{2}\frac{1}{\beta a} \; , \quad n\in\mathbb{Z} \; .
\end{equation}
In particular ${\cal A}$ can be zero.

Proceeding with the minimisation, let us set $x=1/2$ and thus obtain for
the OM-potential
\begin{equation}
   (\kappa_4)^4 U_{OM}
 = \frac{(|h|\beta)^2}{2 d J V}
   e^{-J\beta} \; .
  \label{SimpleOMPotential}
\end{equation}
Notice once more that this comes from the subleading terms as the leading
terms vanish. Because $J$ and $V$ are $R$ dependent, $U_{OM}$ becomes a
function of $R$ which can be minimised with respect to $R$. However, it
is more convenient to minimise with respect to $J$ instead since
alternatively $V$ can be viewed as a function of $J$ once we have fixed
$x=1/2$. The vanishing of the first derivative of $U_{OM}$ with respect
to $J$ leads to the condition\footnote{By $(\hdots)_J$ we denote the
derivative $d(\hdots)/dJ$.}
\begin{equation}
  \frac{V_J}{V}+\frac{1}{J}+\beta = 0 \; .
   \label{OMMini}
\end{equation}
The derivative of the average linear volume with respect to $J$ is
given by
\begin{equation}
  V_J = \left(\frac{J}{3V}-\frac{1}{r_{OM}}
        \left(\frac{6V}{d}\right)^{1/3}
        \right)^{-1} \; ,
  \label{VDerivative}
\end{equation}
where we have defined the flux-parameter
\begin{equation}
   r_{OM}\equiv r\Big(\frac{1}{2}\Big) = r_v-\frac{r_{M5}}{4}
\end{equation}
which controls the ``running'' of $V$ with $R$
\begin{equation}
V=V_1-r_{OM}R
\end{equation}
for the case with the M5 located in the middle of the orbifold-interval.

Let us now solve (\ref{OMMini}) in the moduli-region $J^2\gg V\gg J\gg
1$. By neglecting the $1/J$ against the $\beta$ term and employing
(\ref{VDerivative}), one obtains upon again neglecting $1/\beta$
against $J/3$
\begin{equation}
\frac{1}{r_{OM}}\Big(\frac{6V^4}{d}\Big)^{\frac{1}{3}} = \frac{J}{3} \; .
\end{equation}
Hence we have to constrain $r_{OM}$ to positive values. Since $V\gg J$,
the validity of this equation {\em requires a rather large}
$r_{OM} d^{1/3}\gg 1$. With $J=(6V)^{1/3}(V_1-V)/(r_{OM} d^{1/3})$ it
is then easy to arrive at the final solution which gives the {\em
stabilised values of the moduli}
\begin{equation}
  V = \frac{V_1}{4} \; , \qquad   R = \frac{3V_1}{4r_{OM}} \; .
 \label{OMSolution}
\end{equation}

To actually show that the above solution corresponds to a minimum of
the potential and not just to an extremum, we have to show that the
second derivative of the potential with respect to $R$ is positive.
Because
\begin{equation}
U_{OM,RR}=\frac{d^2J}{dR^2}U_{OM,J}+\Big(\frac{dJ}{dR}\Big)^2
U_{OM,JJ}
\end{equation}
and both $d^2J/dR^2$ and $U_{OM,J}$ are negative, it suffices to show
that $U_{OM,JJ}$ is positive in order to establish a minimum.
Explicitly, the second derivative of the potential is proportional to
\begin{equation}
  U_{OM,JJ} \propto \frac{e^{-J\beta}}{JV}
  \bigg( \Big[ \frac{V_J}{V}+\frac{1}{J}+\beta \Big]^2
        +\frac{1}{J^2}+\Big(\frac{V_J}{V}\Big)^2
        -\frac{V_{JJ}}{V} \bigg)
\end{equation}
The first term in square brackets vanishes at the extremal point by
using the extremality condition (\ref{OMMini}). The remaining terms
are manifestly positive except for the last one containing the second
derivative of $V$. However, with the help of (\ref{OMMini}) it can be
written as
\begin{equation}
  -\frac{V_{JJ}}{V}
  = \frac{1}{3}\Big[\beta+\frac{1}{J}\Big]^2
    \bigg( 1+\Big[\beta+\frac{1}{J}\Big]
             \Big[J+\frac{1}{r_{OM}}\Big(\frac{6V^4}{d}\Big)^{1/3}
             \Big]
    \bigg) \; ,
\end{equation}
which shows that the second derivative of the potential is positive at
its extremal point which therefore represents a minimum of the potential.

We emphasize that this minimum of the potential only occurs because we
have a non-constant CY volume whose running along the orbifold-interval
is caused by the non-trivial G-flux. In contrast a constant CY volume and
thereby an $R$ independent average $V$ would lead to the well-known
runaway-behaviour for (\ref{SimpleOMPotential}).

\section{Properties of the OM Instanton Stabilised Vacuum}
\subsection{G-Fluxes and the Validity of the First Order Approximation}
Let us now check for what values of $r_v$ and $r_{M5}$ we can trust the
obtained solution, i.e.~the first order approximation. Evaluating the
corresponding constraints for the vacuum (\ref{OMSolution}), either
(\ref{Con1}) with (\ref{Con2a}) or (\ref{Con2b}) together
with (\ref{Con2}), both lead to the flux-equality
\begin{equation}
  r_v = r_{M5} \quad \Rightarrow \quad
  r_{OM} = \frac{3}{4}r_v
\end{equation}
which can be used to express the obtained stabilised value for $R$ purely
in terms of visible boundary data
\begin{equation}
  R = \frac{V_1}{r_v} \; .
\end{equation}
Thus the solution saturates the bound $R \le V_1/r_v$ imposed by
(\ref{Con1}) which means that the CY volume $V(x^{11})$ becomes zero at
the location of the M5.

Thus compatibility of the stabilised solution with the first order
approximation gives a precise relationship between the fluxes on the
visible boundary and the M5. Taken together with the anomaly cancellation
constraint (\ref{ACC}), one obtains
\begin{equation}
r_h=0
\end{equation}
and hence the following
relationships
\begin{equation}
\text{tr}F_{(2)}^2 = \frac{1}{2}\text{tr}R^2 \; ,     \qquad
-\big( \text{tr}F_{(1)}^2 - \frac{1}{2}\text{tr}R^2 \big)
= 8\pi^2[\Sigma] \; .
\end{equation}
These lead to a relation between the ``instanton-numbers'' on the two
boundaries
\begin{equation}
 -\int_{CY_3}\omega\wedge\text{tr}F_{(1)}^2
 +\int_{CY_3}\omega\wedge\text{tr}F_{(2)}^2
 = 8\pi^2\int_\Sigma\omega
 = 8\pi^2 Vol(\Sigma) \; ,
\end{equation}
their difference being determined by the G-flux jump coming from the M5.
Notice that the rhs is proportional to
\begin{equation}
  8\pi^2 \epsilon \int_{CY_3} \omega \wedge [\Sigma] \equiv W_G \; ,
\end{equation}
where $W_G$ is the tree-level superpotential generated by the G-flux
of the M5 brane (see e.g.~\cite{LOW},\cite{BG}, and also
\cite{GVW},\cite{HL} for the CY fourfold case). This is what one could have
expected on account of energy-conservation reasoning \cite{GVW}, namely that the
G-flux from the M5 leads to a flux jump which is responsible for
the difference between the boundary G-fluxes. We remark that it was not
necessary to include in (\ref{InitSupo}) this type of superpotential or a
related one stemming from the dimensional reduction of the Chern-Simons
term, $C\wedge G\wedge G$, of eleven-dimensional supergravity for the
following reason. As has been shown in \cite{LOW} they are of higher
order in $\epsilon$ than the leading contributions considered in
(\ref{InitSupo}).

Eventually, we have to verify that the expansion
parameters $\epsilon$ and $\epsilon_R$ stay small. For the above solution
this requires that
\begin{equation}
  \epsilon = 3 2^{1/3} \frac{V_1^{1/3}}{r_{OM}} < 1 \; , \qquad
  \epsilon_R = \sqrt{\frac{\pi}{2}}\frac{2^{5/3}r_{OM}}{3V_1^{5/6}} < 1 \; .
\label{Lat3}
\end{equation}
In particular this implies that $V_1>8\pi\simeq 25.1$ and $r_{OM} >
3(16\pi)^{1/3} \simeq 11.1$. To show that these two constraints
actually do have a common solution, we have plotted in fig.4 and fig.5
in appendix C the two expansion parameters, $\epsilon$ and $\epsilon_R$
in the region $525\le V_1\le 5000$, $80\le r_v \le 250$ with $d = 30$.
The average CY volume chosen is the one appropriate for the OM case
(i.e.~with $x=1/2$).
             \begin{figure}[t]
              \begin{center}
               \begin{picture}(130,100)(0,-5)
                  \LongArrow(5,2)(5,100)
                  \LongArrow(-10,5)(120,5)

                  \Text(-15,50)[]{vis.}
                  \Text(-15,40)[]{bound.}
                  \DashLine(55,5)(55,70){4}
                  \Text(68,45)[]{M5}
                  \DashLine(105,5)(105,70){4}
                  \Text(126,50)[]{hid.}
                  \Text(126,40)[]{bound.}

                  \Line(5,80)(55,5)
                  \Line(55,6)(105,6)

                  \Line(55,3)(55,7)
                  \Line(105,3)(105,7)
                  \Line(3,80)(7,80)

                  \Text(27,100)[]{$V(x^{11})$}
                  \Text(-5,80)[]{$V_1$}
                  \Text(4,-5)[]{$0$}
                  \Text(55,-5)[]{$R\rho/2$}
                  \Text(105,-5)[]{$R\rho$}
                  \Text(134,8)[]{$x^{11}$}

               \end{picture}
               \caption{\it The CY volume dependence on the orbifold coordinate
                        $x^{11}$ in the eleven-dimensional picture which
                        is implied by the stabilised moduli and
                        G-flux values found within the four-dimensional
                        effective description.}
              \end{center}
             \end{figure}
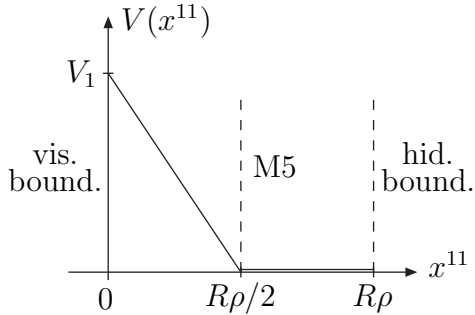

\subsection{The Eleven-Dimensional Picture}
We can also infer to which kind of eleven-dimensional geometry this
flux relation corresponds to. It is easy to see that the obtained
stabilised $V$ and $R$ moduli values together with the equality of the
G-fluxes imply that in the eleven-dimensional picture the variation of
the CY volume (not its average) with the orbifold coordinate $x^{11}$
is as follows (see fig.1)
\begin{equation}
     V(x^{11}) =
     \left\{ \begin{array}{lll}
     V_1-\frac{2}{\rho}r_vx^{11} & \; , &
     0 \le x^{11}< \frac{R\rho}{2} \\
     \qquad 0           & \; , &
     \frac{R\rho}{2}\le x^{11} \le R\rho
     \end{array} \right.
\end{equation}
It might seem bizarre that the eleven-dimensional geometry exhibits a
zero CY volume along an interval. There is however reason to believe that
this is so only in the first order approximation but no longer the case
in a full treatment of heterotic M-theory. To explain this, let us assume
that beyond the first order approximation two features of the stabilised
vacuum remain true. First, the $x\leftrightarrow 1-x$ exchange symmetry
should remain valid since nothing distinguishes one of the OM's against
the other. This means that $x=1/2$ would remain the equilibrium position
of the M5.  Second, let us assume that in addition the equality of the
fluxes on the visible boundary and the M5 remains valid. With these
two assumptions, it is possible to use the result of \cite{CK} to
obtain the eleven-dimensional CY volume behaviour in the full
non-linear treatment\footnote{One has to identify the flux ${\cal
S}_1$ in the notation of \cite{CK} with $r_v/(V_1\rho)$ in the
notation used here. Similarly ${\cal S}_{M5}$ there has to be
identified with $-r_{M5}/(V_1\rho)$ here.}
\begin{equation}
     V(x^{11}) =
     \left\{ \begin{array}{lll}
     \big(1-\frac{r_v}{V_1\rho} x^{11}\big)^2 V_1 & \; , &
     0 \le x^{11}< x_{M5}=\frac{R\rho}{2} \\
     \;\; \big(1-\frac{r_vR}{2V_1}\big)^2 V_1     & \; , &
     x_{M5} \le x^{11} \le R\rho
     \end{array} \right.
\end{equation}
            \begin{figure}[t]
              \begin{center}
               \begin{picture}(130,100)(0,-5)
                  \LongArrow(5,2)(5,100)
                  \LongArrow(-10,5)(120,5)

                  \Text(-15,60)[]{vis.}
                  \Text(-15,50)[]{bound.}
                  \DashLine(55,5)(55,70){4}
                  \Text(68,55)[]{M5}
                  \DashLine(105,5)(105,70){4}
                  \Text(126,60)[]{hid.}
                  \Text(126,50)[]{bound.}

                  \Curve{(5,80)(15,65.75)(25,53)(35,41.75)(45,32)(55,23.75)}
                  \Line(55,23.75)(105,23.75)


                  \Line(55,3)(55,7)
                  \Line(105,3)(105,7)
                  \Line(3,80)(7,80)
                  \Line(3,23.75)(7,23.75)

                  \Text(27,100)[]{$V(x^{11})$}
                  \Text(-5,80)[]{$V_1$}
                  \Text(-15,23.75)[]{$V_1/4$}
                  \Text(4,-5)[]{$0$}
                  \Text(55,-5)[]{$R\rho/2$}
                  \Text(105,-5)[]{$R\rho$}
                  \Text(134,8)[]{$x^{11}$}

               \end{picture}
               \caption{\it The CY volume behaviour which is found beyond
                        leading order under the assumptions that $x=1/2$
                        and $r_v = r_{M5}$ remain true in the full theory.
                        Over the first half of the orbifold-interval the
                        volume varies quadratically and stays constant over
                        the second half. The zero volume interval gets
                        lifted to a positive value $V_1/4$.}
              \end{center}
            \end{figure}
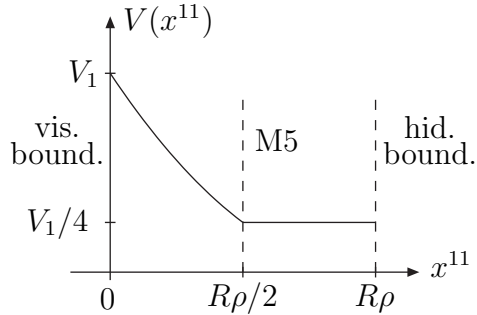
Over the first part of the interval the CY volume varies quadratically
while over the second part it stays constant as a consequence of the
flux-equality. It is interesting now to substitute for $R$ the value
found for the stabilised OM vacuum
\begin{equation}
   R = \frac{V_1}{r_v}
\end{equation}
which gives no longer a vanishing but positive value
$V(x^{11})=V_1/4$ for the second part of the orbifold-interval (see
fig.2).

\subsection{Vacuum Energy}
To illustrate graphically that the obtained extremising solution
(\ref{OMSolution}) actually corresponds to a minimum of the potential
we have plotted the logarithm of the OM potential in fig.3 for the choice of
parameters (which are representative for the orders of magnitude needed
to obey the constraints (\ref{Lat3}))
\begin{equation}
  |h|=\beta=1 \; , \quad V_1=3000 \; , \quad r_{OM}=200 \; , \quad d=30
  \; .
\end{equation}
\begin{figure}[t]
  \begin{center}
  \epsfig{file=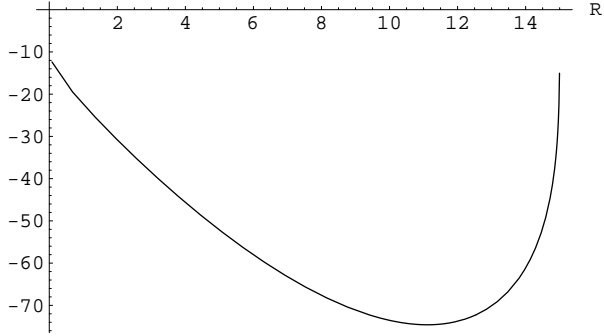,width=8cm,angle=0}
  \caption{\it The logarithm of the OM potential, $\ln((\kappa_4)^4U_{OM})$,
               is depicted as a function
               of the orbifold modulus $R$ for parameters
               $|h|=\beta=1$, $V_1=3000$, $r_{OM}=200$, $d=30$. It
               exhibits a minimum at $R=11$. At $R=15$ the average CY
               volume $V$ vanishes thus leading to the steep increase there.
               The reason for this is that the CY volume becomes negative
               to the right of the minimum and one can strictly trust the
               potential only up to its minimum. The possibility of a
               saddle point at $R=11$ is however excluded since the potential
               exhibits a positive second derivative there.}
  \end{center}
\end{figure}
Indeed, the OM potential exhibits a minimum around $R=11$ in agreement
with (\ref{OMSolution}). Due to the exponential suppression by the factor
$e^{-J\beta}$ the contribution to the vacuum energy can be remarkably
low. Indeed, e.g.~by choosing parameter values like
\begin{equation}
  |h|=\beta=1 \; , \quad V_1=5400 \; , \quad r_{OM}=100 \; , \quad d=30
\; .    \label{SampleVals}
\end{equation}
it is possible to lower this contribution to the vacuum energy to the order of
\begin{equation}
  U_{OM} \simeq 10^{-121}M_{Pl}^4 \simeq \text{meV}^4 \; ,
\end{equation}
which is the observed scale of the cosmological constant. One has to
note, however, that the complete vacuum energy will also comprise the
quantum fluctuations of other fields like the gauge fields for
example. These are not suppressed likewise and therefore one still
faces the cosmological constant problem. To suppress them likewise
another mechanism like e.g.~a suppression by higher-dimensional
warp-factors might be a prospect (see \cite{K2} for a purely
geometrical approach). In the context of the still fictitious
full M-Theory one also has to keep in mind that T-duality can
change the value of the cosmological constant \cite{HW} and thus there is
some arbitrariness in its definition as long as one does not ``fix'' this
duality.

In the last section when we come to the issue of the scale of
supersymmetry-breaking, it will turn out that to achieve $M_{Susy}\simeq$
TeV requires smaller values for $V_1$ and/or larger values for $r_{OM}$
than those given in (\ref{SampleVals}) and thus the vacuum energy
contribution becomes much bigger. For the solution found the OM instanton
contribution to the vacuum energy reads
\begin{equation}
  (\kappa_4)^4 U_{OM} \simeq 2.3 \Big(\frac{|h|\beta}{d^{1/3}}\Big)^2
                      \frac{r_{OM}}{V_1^{7/9}}
                      e^{-0.9\frac{\beta V_1^{4/3}}{r_{OM}d^{1/3}}}
\; .
\end{equation}

\section{Gravitino-Mass and Supersymmetry-Breaking Scale}
We have seen that in the effective four-dimensional description of
heterotic M-theory OM-instantons generically break
supersymmetry (for earlier considerations of supersymmetry-breaking in
heterotic M-theory by gaugino condensation see \cite{NOY}). In order
to determine the supersymmetry-breaking scale $M_{Susy}$, we have to
calculate the F-terms of the respective chiral moduli supermultiplets
and determine their vev's \cite{FGN}. The F-terms are given by
\begin{equation}
  F^i = e^{\frac{K}{2}} D^i W
      \equiv e^{\frac{K}{2}} K^{i\jbar} D_{\jbar} \Wbar \; ,
      \qquad i=S,T,Z \; .
\end{equation}
In terms of them and the generalized K\"ahler-potential $G$, given by
$e^G = e^K |W|^2$, the potential of four-dimensional N=1 supergravity
can be expressed as
\begin{equation}
  (\kappa_4)^4 U = K_{i\jbar}F^i F^{\jbar} - 3 e^G \; .
\end{equation}
The scale $M_{Susy}$ of the supersymmetry-breakdown is given by the
vev of $F^i$ through
\begin{equation}
  M_{Susy}^2 = M_{Pl}^2 |\langle F^i \rangle|
             = M_{Pl}^2 e^{\frac{K}{2}} |D^i W| \; .
\end{equation}

The other interesting quantity related to supersymmetry-breaking is
the value of the gravitino mass $m_{3/2}$ which is given by
\begin{equation}
  m_{3/2}^2 = M_{Pl}^2 e^{\frac{K}{2}} |W|
            = M_{Pl^2}^2 e^{\frac{G}{2}} \; .
\end{equation}
Thus, in order to determine $M_{Susy}$ and $m_{3/2}$, we have to
know $e^{\frac{K}{2}}$, $|W|$ and $|D^i W|$ for the OM
stabilised vacuum examined previously. This is derived in appendix B and
leads to the following expressions
\begin{alignat}{3}
  m_{3/2} &\simeq M_{Pl}\sqrt{|h|}
            \Big(\frac{6e^{-J}}{d V J^3}\Big)^{1/4} \\
  M_{Susy} &\simeq J m_{3/2} = M_{Pl}\sqrt{|h|}
            \Big(\frac{6J}{d V} e^{-J}\Big)^{1/4} \; .
\end{alignat}
We have equal F-terms for $S$, $T$, and $Z$ all giving rise to the same
$M_{Susy}$.

\subsection{Comparison of Vacuum Energy with $\boldsymbol{M_{Susy}}$}
It is interesting to compare the vacuum energy of the OM case with its
supersymmetry breaking scale. We obtained for the vacuum energy
\begin{equation}
U_{OM}^{1/4} \simeq M_{Pl}\sqrt{|h|}
\Big(\frac{6e^{-J}}{dJV}\Big)^{\frac{1}{4}} \; .
\end{equation}
From phenomenological reasoning one would like to have
\begin{equation}
  M_{Susy} \gg U_{OM}^{1/4} \; .
\end{equation}
With the above formula for $M_{Susy}$ this tanslates into
\begin{equation}
  \sqrt{J} \gg 1  \; .
\end{equation}
It is satisfying to see that this is true in the considered region of
moduli space, where $J\gg 1$. However, to become more realistic a huge
value of $J\simeq 10^{30}$ would be needed to bridge the gap between
the observed meV vacuum energy and a TeV supersymmetry breaking
scale. This is however far beyond the values of $J$ considered in this
paper which had to be rather small to guarantee the reliability of the
perturbative formulation of heterotic M-theory.

\subsection{$\boldsymbol{M_{Susy}}$ for the OM Vacua}
Let us finally evaluate $M_{Susy}$ for the OM-instanton vacuum
in terms of the CY data $V_1,d,r_{OM}=3r_v/4$. Using the vacuum given by
(\ref{OMSolution}) we obtain to leading order
\begin{equation}
  M_{Susy} = 2.3 M_{Pl} \sqrt{|h|}
             \bigg( \frac{1}{r_v}\Big( \frac{V_1}{d^4} \Big)^{1/3}
e^{-\frac{1}{r_v}\big(\frac{3V_1^4}{2d}\big)^{1/3}}
\bigg)^{1/4} \; .
\end{equation}
In fig.6 in appendix C we plot $M_{Susy}$ as a function of $V_1$ and
$r_v$ in the region $525\le V_1\le 5000$, $80\le r_v\le 250$ for fixed
values $|h|=1$, $d=30$. As evident from fig.4 and fig.5 (see
appendix C) in this region of parameter space we can trust the
perturbative approach, since both $\epsilon$ and $\epsilon_R$ stay
smaller than one throughout this region and thereby guarantee that
higher order contributions are sufficiently suppressed. From fig.6 it
can be seen that in order to reach the TeV scale with $M_{Susy}$,
rather large values for $r_v$ are required in order to diminish the
huge $V_1$ contribution in the exponent.

\bigskip
\noindent {\large \bf Acknowledgements}\\[2ex]
A.K.~thanks Elias Kiritsis and Nikos Tsamis for valuable comments and
wants to thank the theory group at ENS, Paris, in particular Jean
Iliopoulos, for their friendly hospitality during a stay where part of
this work has been done. The work of A.K.~has been partially supported
by RTN contracts HPRN-CT-2000-00122 and HPRN-CT-2000-00131 and INTAS
grant N 99 1590.
\newpage

\begin{appendix}
\section{K\"ahler-Potential and its Derivatives}
\begin{alignat}{3}
  K &= K_{(S,M5)}+K_{(T)} \\
    &= -\ln \left[ S+\Sbar -\frac{(Z+\Zbar)^2}{\beta(T+\Tbar)} \right]
       -\ln[ \frac{d}{6} (T+\Tbar )^3 ]
       + \ln{\cal O}(\epsilon^2, \epsilon_R^2,\epsilon\epsilon_R) \\
    &= -\ln\left[ \frac{8}{3}d VJ^3
       + {\cal O}(\epsilon^2, \epsilon_R^2,\epsilon\epsilon_R)
           \right]
\end{alignat}
from which it follows that
\begin{equation}
  e^K = \frac{3}{8d VJ^3}
       + {\cal O}(\epsilon^2, \epsilon_R^2,\epsilon\epsilon_R) \; .
\end{equation}

\noindent
First derivatives of $K$ with respect to the moduli:
\begin{alignat}{3}
  K_S &= K_{\Sbar} = -\frac{1}{2V}
                     +{\cal O}(\epsilon^2,\epsilon_R^2,\epsilon\epsilon_R) \\
  K_T &= K_{\Tbar} = -\frac{3}{2J}
                      \left[ 1+\frac{x^2\beta}{3}\frac{J}{V}
                     +{\cal O}(\epsilon^2,\epsilon_R^2,\epsilon\epsilon_R)
                      \right] \\
  K_Z &= K_{\Zbar} = \frac{x}{V}
                     +{\cal O}(\epsilon^2,\epsilon_R^2,\epsilon\epsilon_R)
\end{alignat}

\noindent
Second derivatives:
\begin{alignat}{3}
  K_{S\Sbar} &= \frac{1}{4V^2}
               +{\cal O}(\epsilon^2,\epsilon_R^2,\epsilon\epsilon_R)
                \notag\\
  K_{S\Tbar} &= \frac{x^2\beta}{4 V^2}
               +{\cal O}(\epsilon^2,\epsilon_R^2,\epsilon\epsilon_R)
                \notag\\
  K_{S\Zbar} &= -\frac{x}{2 V^2}
               +{\cal O}(\epsilon^2,\epsilon_R^2,\epsilon\epsilon_R)
                \notag\\
  K_{T\Tbar} &= \frac{3}{4J^2}
                \left[ 1+\frac{2x^2\beta}{3}\frac{J}{V}
               +{\cal O}(\epsilon^2,\epsilon_R^2,\epsilon\epsilon_R)
                \right]
                                                  \label{KahlerMatrix}\\
  K_{T\Zbar} &= -\frac{x}{2VJ}
                 \left[ 1+x^2\beta\frac{J}{V}
               +{\cal O}(\epsilon^2,\epsilon_R^2,\epsilon\epsilon_R)
                 \right] \notag\\
  K_{Z\Zbar} &= \frac{1}{2\beta VJ}
                \left[ 1+2x^2\beta\frac{J}{V}
               +{\cal O}(\epsilon^2,\epsilon_R^2,\epsilon\epsilon_R)
                \right] \notag\; .
\end{alignat}
Note that the second terms in the square brackets are of order
$\epsilon$ and are kept since we are analyzing the potential to
subleading order.

The inverse of the second derivatives K\"ahler-matrix exact to subleading
order obeys
\begin{equation}
  K^{-1}K = 1+{\cal O}(\epsilon^2,\epsilon_R^2,\epsilon\epsilon_R)
\end{equation}
and can be obtained as follows. Let us split the K\"ahler-matrix
(\ref{KahlerMatrix}) into its leading and subleading part
\begin{equation}
  K = K_0 + \epsilon K_1 + {\cal O}(\epsilon^2,\epsilon_R^2,
                                    \epsilon\epsilon_R) \; .
\end{equation}
It is easy to show that $M_0$, the inverse to $K_0$ at leading order
\begin{equation}
  M_0 K_0 = 1+{\cal O}(\epsilon,\epsilon_R)
\end{equation}
is given by
\begin{alignat}{3}
  K^{S\Sbar} &= 4V^2                            \; , \qquad
  K^{S\Tbar} = \frac{4}{3}\beta J^2 x^2             \; , \qquad
  K^{S\Zbar} = 4\beta JVx                \notag \\
  K^{T\Tbar} &= \frac{4}{3} J^2                 \; , \qquad
  K^{T\Zbar} = \frac{4}{3}\beta J^2 x                \; , \qquad\quad
  K^{Z\Zbar} = 2\beta JV  \; ,
\end{alignat}
with the missing entries related to the ones given by $K^{i\jbar} =
K^{j\ibar}$ symmetry. The additional subleading piece, $M_1$,
which completes the inverse K\"ahler-matrix at subleading order
\begin{equation}
 K^{-1} = M_0 + \epsilon M_1
          +{\cal O}(\epsilon^2,\epsilon_R^2,\epsilon\epsilon_R)
\end{equation}
is given by
\begin{equation}
 M_1 \equiv -(M_0 K_1 + \Delta) M_0 \; ,
\end{equation}
where $\Delta$ measures the deviation of $M_0K_0$ from the identity at
subleading order
\begin{equation}
  \epsilon \Delta = M_0 K_0 - 1
                  +{\cal O}(\epsilon^2,\epsilon_R^2,\epsilon\epsilon_R) \; .
\end{equation}
Following these steps gives us finally the inverse K\"ahler-matrix,
$K^{-1}$, correct up to subleading order
\begin{alignat}{3}
  K^{S\Sbar} &= 4V^2\left( 1+2x^2\beta\frac{J}{V} \right)
                                                       \; , \qquad\qquad\qquad
  &&K^{S\Tbar} = \frac{4}{3}\beta J^2 x^2                \; , \notag \\
  K^{S\Zbar} &= 4\beta JVx
               \left( 1+\frac{1}{3}x^2\beta\frac{J}{V} \right)
                                                       \; , \qquad
  &&K^{T\Tbar} = \frac{4}{3} J^2                         \; ,        \\
  K^{T\Zbar} &= \frac{4}{3} \beta J^2 x                \; , \qquad\qquad\qquad
                                                            \qquad\qquad\qquad
  &&K^{Z\Zbar} = 2\beta JV
               \left( 1+\frac{2}{3}x^2\beta\frac{J}{V} \right)
                                                            \notag
\end{alignat}
Again the symmetry $K^{i\jbar} = K^{j\ibar}$ gives the remaining
matrix entries.

\section{Technical Details for Deriving $\boldsymbol{M_{Susy}}$ and
         $\boldsymbol{m_{3/2}}$}
We will present in this appendix those expressions which
are needed for the computation of the supersymmetry-breaking scale
$M_{Susy}$ and the gravitino mass $m_{3/2}$.

From appendix A we see that the exponential involving the
K\"ahler-potential is given by
\begin{equation}
  e^{\frac{K}{2}} = \Big(\frac{3}{8dVJ^3}\Big)^{\frac{1}{2}} \; .
\end{equation}

Next, we have to calculate the modulus of the OM superpotential
\begin{equation}
  W = h\left( e^{-Z}+e^{Z-\beta T} \right)
\end{equation}
which turns out to be
\begin{equation}
  |W| = |h| \Big( e^{-2J\beta x}+e^{-2J\beta(1-x)}
                +2e^{-\beta J}\cos(2\alpha-\beta\chi)
            \Big)^{\frac{1}{2}}
\end{equation}
This has to be evaluated for the OM vacuum derived in the main text. For
this we have to use the axion minimisation condition (\ref{Lat1})
together with $n$ even and M5 position modulus $x=1/2$. This leads to the
vacuum expression
\begin{equation}
|W| = 2|h|e^{-\frac{J}{2}} \; .
\end{equation}

The last ingredient is the absolute value
of the K\"ahler-covariant derivatives $|D^i W|$. Let us start from the
derivatives with lower indices first. Their leading
orders\footnote{These are sufficient to determine $|D^iW|$ including
all $JV$ and $J^2$ contributions which give the leading expressions
for $M_{Susy}$ and $m_{3/2}$.} are given by
\begin{alignat}{3}
  D_SW &= -\frac{h}{2V} ( e^{-Z}+e^{Z-\beta T} )  \\
  D_TW &= -\beta h e^{Z-\beta T} \\
  D_ZW &= h ( -e^{-Z}+e^{Z-\beta T} )  \; .
\end{alignat}
The next step is to calculate from these the upper-index derivatives
$D^iW=K^{i\jbar}D_{\jbar}\Wbar$. The general structure of the $D^i W$ can
be parameterised as ($i=S,T,Z$)
\begin{equation}
  D^i W =  A_i \hbarr e^{-\Zbar}
          +B_i \hbarr e^{\Zbar-\beta \Tbar}  \; ,
\end{equation}
where the specific coefficients read
\begin{alignat}{3}
  A_S &= -4JV\beta x\Big(1+\frac{\beta x^2}{3}\frac{J}{V}\Big) \; , \quad
  B_S = -A_S-\frac{4}{3}\beta^2 x^2J^2 \; ,  \notag \\
  A_T &= -\frac{4}{3}\beta xJ^2 \; , \quad
  B_T = -\frac{4}{3}\beta(1-x)J^2 \; , \notag \\
  A_Z &= -2\beta JV\Big(1+\frac{2}{3}\beta x^2\frac{J}{V}\Big)\; ,\quad
  B_Z = -A_Z-\frac{4}{3}\beta^2 x J^2 \; .
\end{alignat}
Its absolute value can then be figured out to be
\begin{equation}
 |D^iW| =  |h| \Big( A_i^2 e^{-2J\beta x}+B_i^2 e^{-2J\beta(1-x)}
                  +2A_iB_ie^{-\beta J}\cos(2\alpha-\beta\chi)
               \Big)^{\frac{1}{2}}  \; .
\end{equation}
Again, to evaluate this expression for the OM vacuum, we use the axion
minimisation condition (\ref{Lat1}) together with $n$ even and $x=1/2$
which gives
\begin{equation}
  |D^iW| = |h||A_i+B_i|e^{-\frac{J}{2}} \; .
\end{equation}
Specifically, this leads to the following vacuum expressions
\begin{equation}
  |D^SW| = \frac{\beta}{4}|D^TW| = \frac{1}{2}|D^ZW|
  = \frac{\beta^2}{3}|h|J^2e^{-\frac{J}{2}} \; .
\end{equation}
Hence we obtain the succinct result
\begin{equation}
  |D^iW| \simeq |h|J^2 e^{-\frac{J}{2}} \; .
\end{equation}

\section{Plots}
\begin{figure}[h]
  \begin{center}
  \epsfig{file=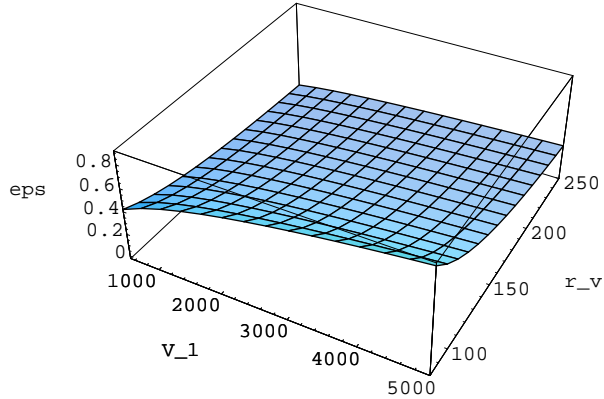,width=8cm,angle=0}
  \caption{\it The figure shows that $\epsilon$ stays smaller than one
               in the $(V_1,r_v)$ parameter region given by
               $525 \le V_1\le 5000$, $80\le r_v\le 250$ and
               $d = 30$.}
  \end{center}
\end{figure}

\begin{figure}[p]
  \begin{center}
  \epsfig{file=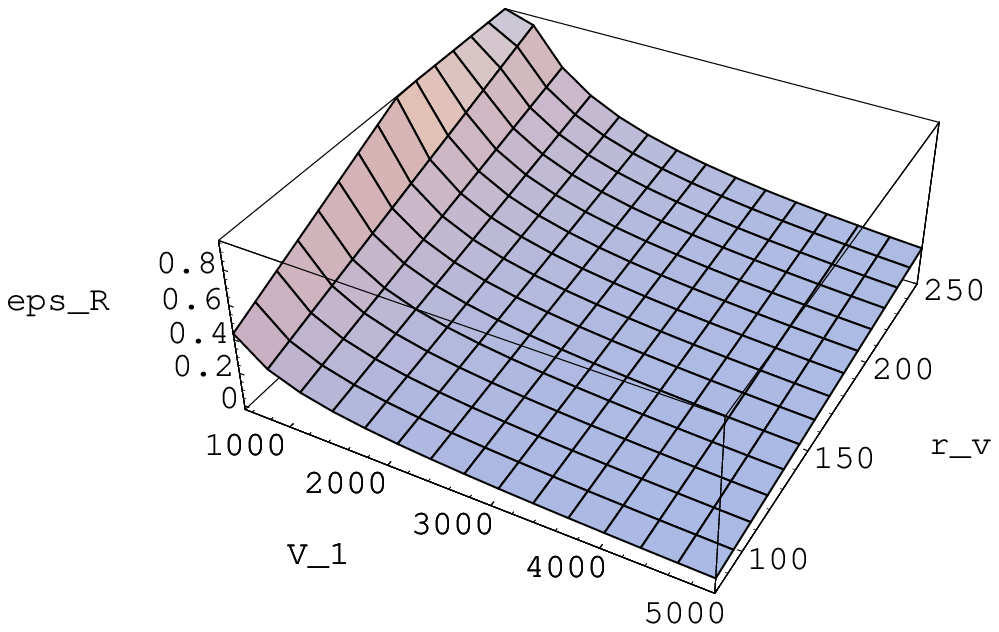,width=8cm,angle=0}
  \caption{\it The figure shows that also $\epsilon_R$ stays smaller than one
               in the same $(V_1,r_v)$ parameter region as considered in
               the previous figure.}
  \end{center}

  \begin{center}
  \epsfig{file=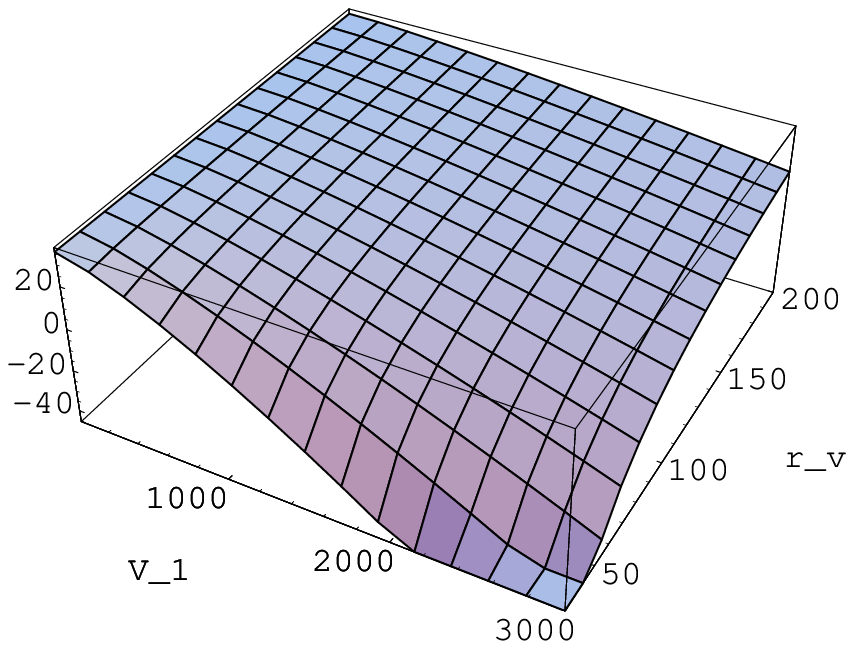,width=8cm,angle=0}
  \caption{{\it The figure shows the dependence of } $\ln(M_{Susy}/\text{TeV})$
          {\it on the two visible boundary CY volume and flux parameters
          $V_1$ and $r_v$. The remaining parameters are set to
          the values $|h|=1$, $\beta =1$, $d = 30$.}}
  \end{center}
\end{figure}

\end{appendix}
\newpage

 \newcommand{\zpc}[3]{{\sl Z. Phys.} {\bf C\,#1} (#2) #3}
 \newcommand{\npb}[3]{{\sl Nucl. Phys.} {\bf B\,#1} (#2) #3}
 \newcommand{\plb}[3]{{\sl Phys. Lett.} {\bf B\,#1} (#2) #3}
 \newcommand{\prd}[3]{{\sl Phys. Rev.} {\bf D\,#1} (#2) #3}
 \newcommand{\prb}[3]{{\sl Phys. Rev.} {\bf B\,#1} (#2) #3}
 \newcommand{\pr}[3]{{\sl Phys. Rev.} {\bf #1} (#2) #3}
 \newcommand{\prl}[3]{{\sl Phys. Rev. Lett.} {\bf #1} (#2) #3}
 \newcommand{\jhep}[3]{{\sl JHEP} {\bf #1} (#2) #3}
 \newcommand{\cqg}[3]{{\sl Class. Quant. Grav.} {\bf #1} (#2) #3}
 \newcommand{\prep}[3]{{\sl Phys. Rep.} {\bf #1} (#2) #3}
 \newcommand{\fp}[3]{{\sl Fortschr. Phys.} {\bf #1} (#2) #3}
 \newcommand{\nc}[3]{{\sl Nuovo Cimento} {\bf #1} (#2) #3}
 \newcommand{\nca}[3]{{\sl Nuovo Cimento} {\bf A\,#1} (#2) #3}
 \newcommand{\lnc}[3]{{\sl Lett. Nuovo Cimento} {\bf #1} (#2) #3}
 \newcommand{\ijmpa}[3]{{\sl Int. J. Mod. Phys.} {\bf A\,#1} (#2) #3}
 \newcommand{\rmp}[3]{{\sl Rev. Mod. Phys.} {\bf #1} (#2) #3}
 \newcommand{\ptp}[3]{{\sl Prog. Theor. Phys.} {\bf #1} (#2) #3}
 \newcommand{\sjnp}[3]{{\sl Sov. J. Nucl. Phys.} {\bf #1} (#2) #3}
 \newcommand{\sjpn}[3]{{\sl Sov. J. Particles \& Nuclei} {\bf #1} (#2) #3}
 \newcommand{\splir}[3]{{\sl Sov. Phys. Leb. Inst. Rep.} {\bf #1} (#2) #3}
 \newcommand{\tmf}[3]{{\sl Teor. Mat. Fiz.} {\bf #1} (#2) #3}
 \newcommand{\jcp}[3]{{\sl J. Comp. Phys.} {\bf #1} (#2) #3}
 \newcommand{\cpc}[3]{{\sl Comp. Phys. Commun.} {\bf #1} (#2) #3}
 \newcommand{\mpla}[3]{{\sl Mod. Phys. Lett.} {\bf A\,#1} (#2) #3}
 \newcommand{\cmp}[3]{{\sl Comm. Math. Phys.} {\bf #1} (#2) #3}
 \newcommand{\jmp}[3]{{\sl J. Math. Phys.} {\bf #1} (#2) #3}
 \newcommand{\pa}[3]{{\sl Physica} {\bf A\,#1} (#2) #3}
 \newcommand{\nim}[3]{{\sl Nucl. Instr. Meth.} {\bf #1} (#2) #3}
 \newcommand{\el}[3]{{\sl Europhysics Letters} {\bf #1} (#2) #3}
 \newcommand{\aop}[3]{{\sl Ann. of Phys.} {\bf #1} (#2) #3}
 \newcommand{\jetp}[3]{{\sl JETP} {\bf #1} (#2) #3}
 \newcommand{\jetpl}[3]{{\sl JETP Lett.} {\bf #1} (#2) #3}
 \newcommand{\acpp}[3]{{\sl Acta Physica Polonica} {\bf #1} (#2) #3}
 \newcommand{\sci}[3]{{\sl Science} {\bf #1} (#2) #3}
 \newcommand{\vj}[4]{{\sl #1~}{\bf #2} (#3) #4}
 \newcommand{\ej}[3]{{\bf #1} (#2) #3}
 \newcommand{\vjs}[2]{{\sl #1~}{\bf #2}}
 \newcommand{\hepph}[1]{{\sl hep--ph/}{#1}}
 \newcommand{\desy}[1]{{\sl DESY-Report~}{#1}}

\bibliographystyle{plain}

\end{document}